\documentclass[fleqn,usenatbib]{mnras}

\usepackage{relsize}
\usepackage{placeins}

\usepackage[T1]{fontenc}
\usepackage{ae,aecompl}
\usepackage{booktabs}

\usepackage[normalem]{ulem}

\DeclareRobustCommand{\VAN}[3]{#2}
\let\VANthebibliography\thebibliography
\def\thebibliography{\DeclareRobustCommand{\VAN}[3]{##3}\VANthebibliography}

\usepackage{graphicx}	%
\usepackage{amsmath}	%
\usepackage{amssymb}	%
\usepackage{hyperref}

\newcommand{\sacf}{S-ACF}
\newcommand{\kepler}{\textit{Kepler}}
\newcommand{\kicID}{KIC 5110407}

\title[\sacf]{\sacf: A selective estimator for the autocorrelation function of irregularly sampled time series}

\author[L. T. Kreutzer et al.]{
Lars T. Kreutzer,$^{1,2,3}$\thanks{E-mail: lars.kreutzer@aei.mpg.de}
Edward Gillen,$^{4,2}$\thanks{Winton Fellow}
Joshua T. Briegal,$^{2}$
Didier Queloz$^{2}$
\\
$^{1}$Department of Applied Mathematics and Theoretical Physics, University of Cambridge, Cambridge, CB3 OWA, UK\\
$^{2}$Astrophysics Group, Cavendish Laboratory, J.J. Thomson Avenue, Cambridge CB3 0HE, UK.\\
$^{3}$Max Planck Institute for Gravitational Physics (Albert Einstein Institute), Am M\"uhlenberg 1, 14476 Golm, Germany\\
$^{4}$Astronomy Unit, Queen Mary University of London, Mile End Road, London E1 4NS, UK
}

\date{Accepted XXX. Received YYY; in original form ZZZ}

\pubyear{2023}
\usepackage{newtxtext,newtxmath}

\begin{document}
\label{firstpage}
\pagerange{\pageref{firstpage}--\pageref{lastpage}}
\maketitle

\begin{abstract}
We present a generalised estimator for the autocorrelation function, \sacf{}, which is an extended version of the standard estimator of the autocorrelation function (ACF). \sacf{} is a versatile definition that can robustly and efficiently extract periodicity and signal shape information from a time series, independent of the time sampling and with minimal assumptions about the underlying process. Calculating the autocorrelation of irregularly sampled time series becomes possible by generalising the lag of the standard estimator of the ACF to a real parameter and introducing the notion of selection and weight functions. We show that the \sacf{} reduces to the standard ACF estimator for regularly sampled time series. 
Using a large number of synthetic time series we demonstrate that the performance of the \sacf{} is as good or better than commonly used Gaussian and rectangular kernel estimators, and is comparable to a combination of interpolation and the standard estimator.
We apply the \sacf{} to astrophysical data by extracting rotation periods for the spotted star \kicID{}, and compare our results to Gaussian process (GP) regression and Lomb-Scargle (LS) periodograms. We find that the \sacf{} periods typically agree better with those from GP regression than from LS periodograms, especially in cases where there is evolution in the signal shape.
The \sacf{} has a wide range of potential applications and should be useful in quantitative science disciplines where irregularly sampled time series occur. A Python implementation of the \sacf{} is available under the MIT license.
\end{abstract}


\section{Introduction}

Time series are ubiquitous throughout the experimental sciences and give insight into the temporal evolution of systems and their underlying processes. 
Time series in astrophysics, for example, have been instrumental in our understanding of stellar and planetary systems: stellar light and radial velocity curves yield information about the temporal evolution of processes on the stellar surface, from the longitudinal inhomogeneity of starspot distributions and magnetic field mechanisms, to the presence of orbiting bodies and material.

Historically, detecting periodicity in time series has focused on either Fourier decomposition (for regularly sampled data) or fitting sinusoidal models (for irregularly sampled data). 
An example of the former is the Fast Fourier Transform \citep[FFT;][]{Cooley69}, 
and examples of the latter are the standard, modified and Bayesian Lomb-Scargle periodograms \citep{Scargle82,Zechmeister09,Mortier17}. 
While the Lomb-Scargle method can be used for arbitrary samplings, the accuracy of the estimated periods can be limited for quasi-periodic processes and evolving periodic signals due to the inherent assumption that the process be well-described by a pure sine wave of fixed period. 
Similar issues affect methods based on phase folding and then minimising the variance or entropy of the data, as they also rely on strict periodicity and negligible phase evolution \citep[e.g.][]{Stellingwerf11,Graham13,Graham13a}. 
More recently, flexible machine learning methods, applicable to both regular and irregular time series, have been used to describe quasi-periodic variations in stellar light curves \citep[e.g.][]{Angus18}.

The above approaches share the same basic principle: they all fit a model to the data to determine whether periodicity is present. The concept of autocorrelation, i.e. correlating the data with itself, is a distinct `model-free' approach that uses only the time series data to extract periodicity \citep[e.g.][]{Shumway17}. The autocorrelation function (ACF) is a powerful definition and a reliable method to obtain information from any regularly sampled time series, as it can capture both strictly periodic and quasi-periodic processes. It has been widely used on space-based photometric data given the regular sampling available \citep[e.g.][]{McQuillan13,McQuillan14}, as well as on solar data \citep{Morris19} for the same reason. However, the requirement of the standard ACF estimator for regularly sampled data can be a limiting factor in its broader application, e.g. for ground-based photometric data. 

Previous studies have attempted to address this problem by generalising the standard ACF estimator to irregularly sampled data. A number of these methods create an approximate regularly sampled time series in order to apply the standard autocorrelation estimator enhanced with rules on which terms to discard in the series. The method proposed in \citet{1975Lukatskaia} assumes that the irregular sampling arises from missing data points in a regularly sampled series, and further assumes that statistical properties of the missing data are the same as the observed data. It is therefore possible to calculate an autocorrelation estimator using only data points which fall on to this regular sampling, with the caveat that the time series must be much longer than the variability period of the signal of interest. The method from \citet{Andronov05} interpolates onto a regular sampling grid using a smoothing function.
These methods can work well if the sampling is almost regular and only a subset of values are missing from a regularly sampled time series. 
The Discrete Autocorrelation Function, as proposed in \citet{Edelson88}, relies on binning values in time intervals to account for missing overlaps. A similar method was also proposed in \citet{Mayo1974}.

As an extension of the binning proposed by \citet{Edelson88}, and drawing from the general kernel based methods proposed by \citet{Hall1994Kernel}, \citet{Stoica2006Spectral} and \citet{Bjornstad2001} both use a kernel to weight the product of observations according to the difference between the observation interval and the desired lag bin centre. This is also known as `fuzzy slotting'. The kernels proposed are smooth density functions that tend to zero as lag increases or decreases from the desired lag subject to a characteristic width parameter. \citet{Stoica2006Spectral} propose a sinc function, demonstrating the efficacy of this weighting on examples from over-the-internet temperature data, pulsar time-of-arrival measurements and ice core CO$_2$ measurements. \citet{Bjornstad2001} use a Gaussian kernel in the context of estimating a spatial autocorrelation for sparse ecological population data.
A comparison of a number of correlation analysis techniques for irregularly sampled time series (linear interpolation, Lomb-Scargle periodogram, correlation slotting and several kernel based methods), in a geoscientific context, can be found in \citet{Rehfeld2011NPG}. 
These authors find that while all methods investigated lead to consistent results for time series with a relatively constant sampling density, the kernel based methods perform better for highly irregular time series.

Related to the problem of finding the autocorrelation function of irregularly sampled time series is the problem of finding the power spectral density (PSD), as the Fourier transform of the power spectrum of a (stochastic) time series is equivalent to the ACF \citep[see e.g.][]{Scargle1989III,Merrifield1994PSD}.

In this work we present a different generalisation of the standard estimator of the autocorrelation function, which we name the selective autocorrelation function estimator, or \sacf{}. The \sacf{} is an extended and generalised version of the standard estimator, which is applicable to both regularly and irregularly sampled time series, without making any assumptions about the time sampling or the statistical properties of the discrete time series and only assuming smoothness of the underlying process. In particular, there are no assumptions about regularity in the sampling of the time series.

The \sacf{} method, presented in this publication, has previously been used under the name G-ACF in the publications \citet{Gillen20a} and \citet{Briegal22}.

A Python implementation of the \sacf{} is available under the MIT license at \href{https://www.github.com/joshbriegal/sacf/}{github.com/joshbriegal/sacf/}.

In Section \ref{sec:basic-def} we define our mathematical notation and the standard estimator of the autocorrelation function. We then present the \sacf{} and discuss its main properties in Section \ref{sec:GACF}. In Section \ref{sec:application} we show that the \sacf{} performs accurately on both synthetic and real data. We investigate the effect of different time samplings with both regularly and irregularly sampled time series and find that the \sacf{} produces comparable estimates of the autocorrelation function and corresponding period estimates. 
We conclude in Section \ref{sec:conclusions}.

\section{Basic definitions}
\label{sec:basic-def}

We start with some elementary definitions concerning time series and the standard estimator of the autocorrelation function in order to clarify our vocabulary. We adopt the symbol $:=$ to indicate a definition. An overview of the notation can be found in Appendix \ref{notation}.

\subsection{Time series}

We define a \textit{time series} $X_{I}(t)$ to be a finite ordered set
\begin{equation}
X_I(t) := \{(X_i,t_i)\in\mathbb{R}\times\mathbb{R}^{+}|i\in I\subset\mathbb{N},\quad (t_{i+1}-t_i)>0\: \forall i\in I \}
\end{equation}
with $I\subset\mathbb{N}$ being a finite index set, which we can choose to be $I=\{0,1,2,\dots,i_\text{max}\}$.
Furthermore we will refer to the set $T_I := \{t_i|i\in I\}$ as the set of \textit{time labels} and to the set $X_I := \{X_i|i\in I\}$ as the set of \textit{time series values}. %

It can be useful to think of a time series $X_I(t)$ as a discrete sampling of a continuous process $X(t)$. The notation $X_i = X(t_i)$ will be used. Hence we define a time series to be \textit{regularly sampled} if there exists a \textit{sampling constant} $\Delta t > 0,$ such that $t_k = t_0 + k \cdot \Delta t \quad \forall k\in I$, else we call the time series \textit{irregularly sampled}. %

\subsection{Standard estimator of the autocorrelation function (ACF)}

The \textit{standard estimator of the autocorrelation function} \citep[ACF; e.g. as described in][]{Scargle1989III,Shumway17} relies on the self-similarity of the underlying process of the time series 
and can be applied to all regularly sampled time series to obtain information such as periodicity and signal shapes. We define the standard estimator of the autocorrelation function of a regularly sampled time series $X_I(t)$ as the function 
\begin{align}
\label{ACF_rho}
  \rho & : \{0,1,\dots,i_{\text{max}}\} \to  [-1,1]
\end{align} 
\begin{align}
\label{ACF}
  \rho &(k) := \frac{1}{N} \sum\limits_{i=0}^{i_{\text{max}}-k} (X_i - \langle X_I\rangle)\times(X_{i+k} - \langle X_I\rangle)
\end{align} 
where $\langle X_I\rangle$ denotes the mean of the time series values and the normalisation $N$ is the total sum of squares $N := \sum\limits_{i\in I} \left(X_i - \langle X_I\rangle\right)^2$. The choice of this normalisation implies that $\rho(0) \equiv 1$, i.e. a time series is maximally similar to itself when there is no lag. The argument $k$ is referred to as the \textit{lag}. %

Equation \ref{ACF} is the standard estimator of the (true) autocorrelation function in the case of regularly sampled and finite time series. In the following we will use the abbreviation ACF to mean this standard estimator.

While this is a standard way to introduce the ACF, it can be useful to think of the ACF as a function with a time domain instead of an integer lag domain. 
We can make this domain modification explicit by multiplying the argument by the sampling constant $\Delta t$:
\begin{equation}
\rho(k\Delta t)  : \{0, \Delta t,\dots,\Delta t\cdot i_{\text{max}}\} \to  [-1,1].
\end{equation}
These descriptions are equivalent, but the latter view is more useful for the forthcoming discussion.
Specifically, it is clear that the standard estimator is only directly applicable to time series where the sampling is regular.

As discussed in the Introduction, various efforts have been made to interpolate or infer missing data to recover a regularly sampled time series and allow the use of the standard estimator. The accuracy of such efforts depends on the length of the gaps or irregularities in the time series sampling compared to the scale of structures in the signal. If the time series has large temporal gaps compared to the scale of the underlying process it is in general very difficult to restore the missing information using interpolation. Here, we seek to develop a new method to obtain information from arbitrary time series, regardless of their sampling, which is applicable to all time series. The approach presented in this work is a generalisation of the standard estimator, based on the fact that the standard estimator is already applicable to all regularly sampled time series irrespective of the underlying process. The key step is to formalise the time label dependence of the definition explicitly.

\section{Selective autocorrelation function estimator (\sacf)}
\label{sec:GACF}

We begin generalising the definition of the standard estimator for time series of arbitrary sampling by introducing two functions, the \textit{selection function} $\hat{S}$ and the \textit{weight function} $\hat{W}$, as well as generalising the notion of the lag to a \textit{generalised lag} ${\hat{k} \in [0\, ,\left(\max(T_I)-\min(T_I)\right)] }$.  %

We define the selective autocorrelation function estimator, for a time series of any sampling, to be the function $\hat{\rho}(\hat{k};\, \hat{W},\hat{S})$ which, restricted to the generalised 
lag $\hat{k}$, is a function of the form
 \begin{equation}
  \hat{\rho}(\hat{k}): [0,\left(\max(T_I)-\min(T_I)\right)] \to  [-1,1]. 
 \end{equation}

A possible generalised definition is then given by
\begin{align}
\label{G-ACF}
  \hat{\rho}&\left(\hat{k}; \hat{W},\hat{S}\right) := \nonumber\\
  & \frac{1}{N} \mathlarger{\mathlarger{\sum}}\limits_{\substack{i\in I\\ t_i+\hat{k} \leq \max(T_I) }} \Bigg[  \left(X(t_i) - \langle X_I\rangle\right) \times \left(X(\hat{S}(t_i+\hat{k})) - \langle X_I\rangle\right)
  \nonumber \\
 & \qquad\qquad\qquad\quad\times\,\hat{W}\left(\left|\hat{S}\left(t_i + \hat{k}\right)-\left(t_i + \hat{k}\right) \right|\right)\Bigg].
\end{align}

\noindent Where $N := \sum\limits_{i\in I} \left(X_i - \langle X_I\rangle\right)^2$ denotes the total sum of squares and $\langle X_I\rangle$ is the mean of the time series values set. The general form of the \sacf{} is very similar to that of the standard estimator (Equation \ref{ACF}). The \sacf{} differs from the standard estimator by the explicit inclusion of the selection function in the second factor, the restriction on the sum, the generalised lag and an additional third factor given by the weight function. In the following sub-sections we discuss the three new components: the generalised lag, the selection function and the weight function.

For regularly sampled time series, we want the \sacf{} to reduce to the standard estimator when restricting the generalised lag to multiples of the sampling constant. 
A full proof and detailed explanation of this property is given in Appendix \ref{proofofthereduction}.
This reduction to the standard estimator is one of the core requirements of our generalisation and ensures that the \sacf{} and the standard estimator are equivalent for regularly sampled time series. We note that this requirement motivates some of the restrictions that we impose on the selection and weight functions.

In the case of the ACF, the equation $\rho(0)=1$ tells us that -- trivially -- if we do not shift the time series the correlation is perfect. 
The property $\hat{\rho}(0)=1$ should also hold for the \sacf{}, which we prove in Appendix \ref{Proof-of-property-2}.

\subsection{The generalised lag \texorpdfstring{$\hat{k}$}{k}}

As shown in Equation \ref{G-ACF}, the generalised lag ${\hat{k} \in [0\, ,\left(\max(T_I)-\min(T_I)\right)] }$ can now take any value within an interval in time instead of solely integer values. 
This is natural since in the most general case of irregular sampling there is no preferred time scale -- i.e. there exists no sampling constant -- and thus we can allow the lag to be a continuous variable.

Even though the \sacf{} is a well defined function for any lag, it cannot contain meaningful information at a higher resolution than the time series itself, and we suggest setting the time-resolution of the generalised lag to values no smaller than the minimal difference between two neighbouring time labels $\delta\hat{k}\geq \min(t_{i+1}-t_i)$ for ${\{t_i,t_{i+1}\}\in T_I}$.

The condition $t_i+\hat{k} \leq \max(T_I)$ on the sum is the generalisation of the upper limit ${i_{\text{max}}-k}$ of the sum in the ACF definition (Equation \ref{ACF}). This bound on the (generalised) lag enforces again that the maximum shifting of the process along itself is equal to the temporal length of the time series and thus when the first time label is matched up with the last time label. 

\begin{figure}
  \includegraphics[width=\linewidth]{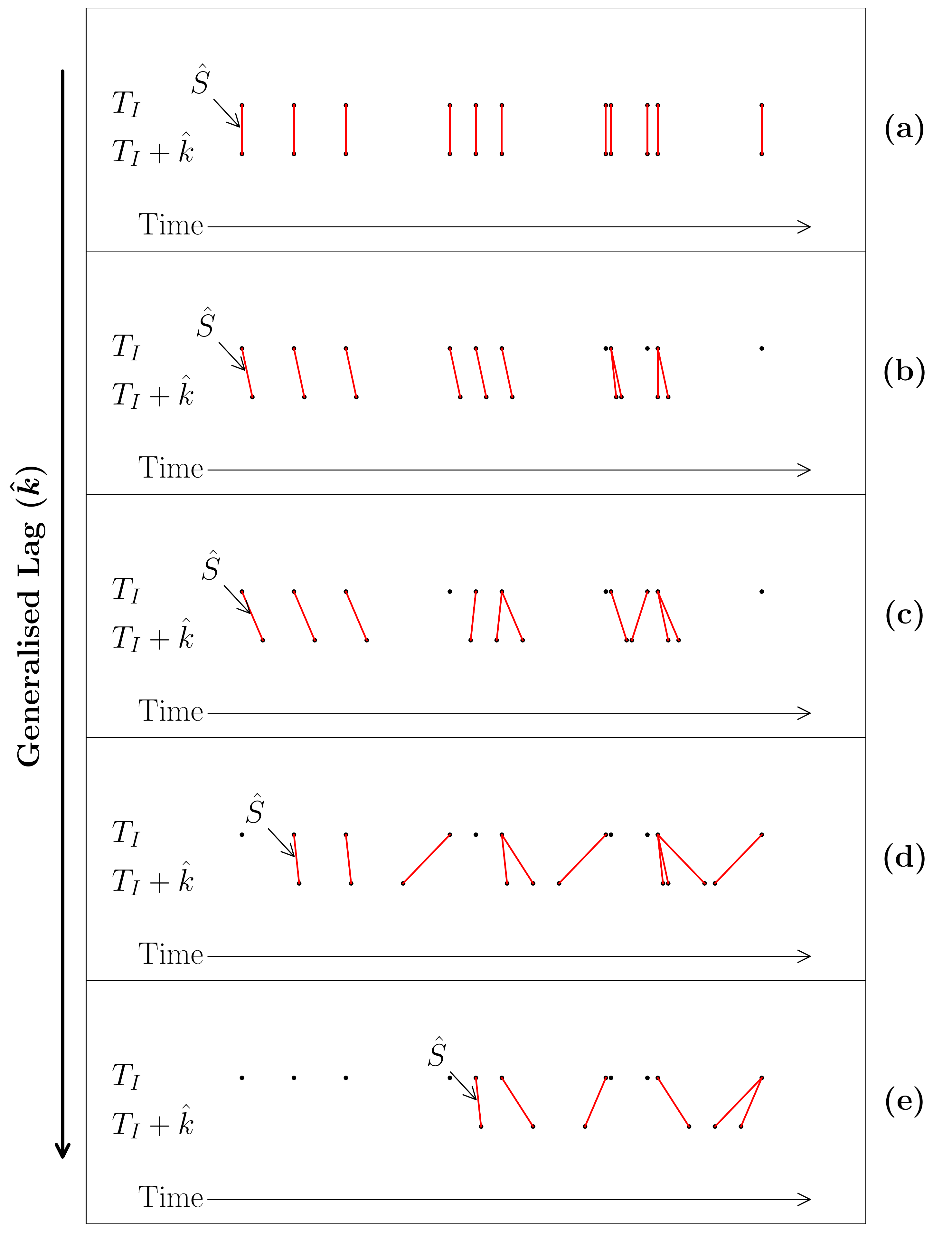}
	\caption{Each graphic (a) to (e) shows a set of time labels on the real axis and below them the same set of time labels shifted by a real generalised lag $\hat k$. The red lines indicate how the selection function $\hat S$ matches the shifted time labels to the original set of time labels above by choosing the closest time label from the set of time labels $T_I$. The generalised lag increases from panel (a) to (e), which corresponds to the lower set of labels `shifting' to the right. A supplementary animated version of this figure is available on the journal website.}
	\label{f:sliding_timesteps} 
\end{figure}

\subsection{The selection function \texorpdfstring{$\hat{S}$}{S}}
\label{selectionfunction}

The selection function is arguably the most important part of the \sacf{} definition (Equation \ref{G-ACF}), as it deals with the irregular sampling issue that is at the core of the problem considered in this work.
We define a selection function $\hat{S}$ to be a function ${\hat{S} : \mathbb{R}^+ \to T_I}$ that projects an arbitrary point in time onto the set of available time labels, thus selecting a specific time label for each point in time.
There are many sensible functions that one could choose to accomplish this, however a natural selection function is the one that, for each point in time, selects the \textit{closest allowed time label} (see Figure \ref{f:sliding_timesteps} for an illustration of this function). In the case that two time labels are equally close to the argument one can employ the convention of always choosing the smaller or larger value, or randomise the decision in any practical application of the \sacf{}. 

{ \color{black}
In order for this selection function to be justified we have to make the assumption that the process underlying the discrete time series has some degree of ``smoothness'' in between the time labels.

A key difference of this selection function, compared to the kernel based methods described in the Introduction, see e.g. \citet{Rehfeld2011NPG}, is that the selection function does not have a kernel with a fixed width. The distance in lag between a time label and the selected shifted time label can be as large as half the size of the largest gap in the time series sampling, as can be seen in Figure \ref{f:sliding_timesteps}. 
}

A possible alternative definition of the selection function would be to find the closest time label for the first shifted time label and then pair up all subsequent labels instead of finding the closest time label for each shifted label individually. While this definition reduces the computational complexity, it did not produce as accurate a reconstruction of the standard ACF as taking the closest time label for each individual time label when tested on synthetic time series.

\begin{figure}
  \includegraphics[width=\linewidth]{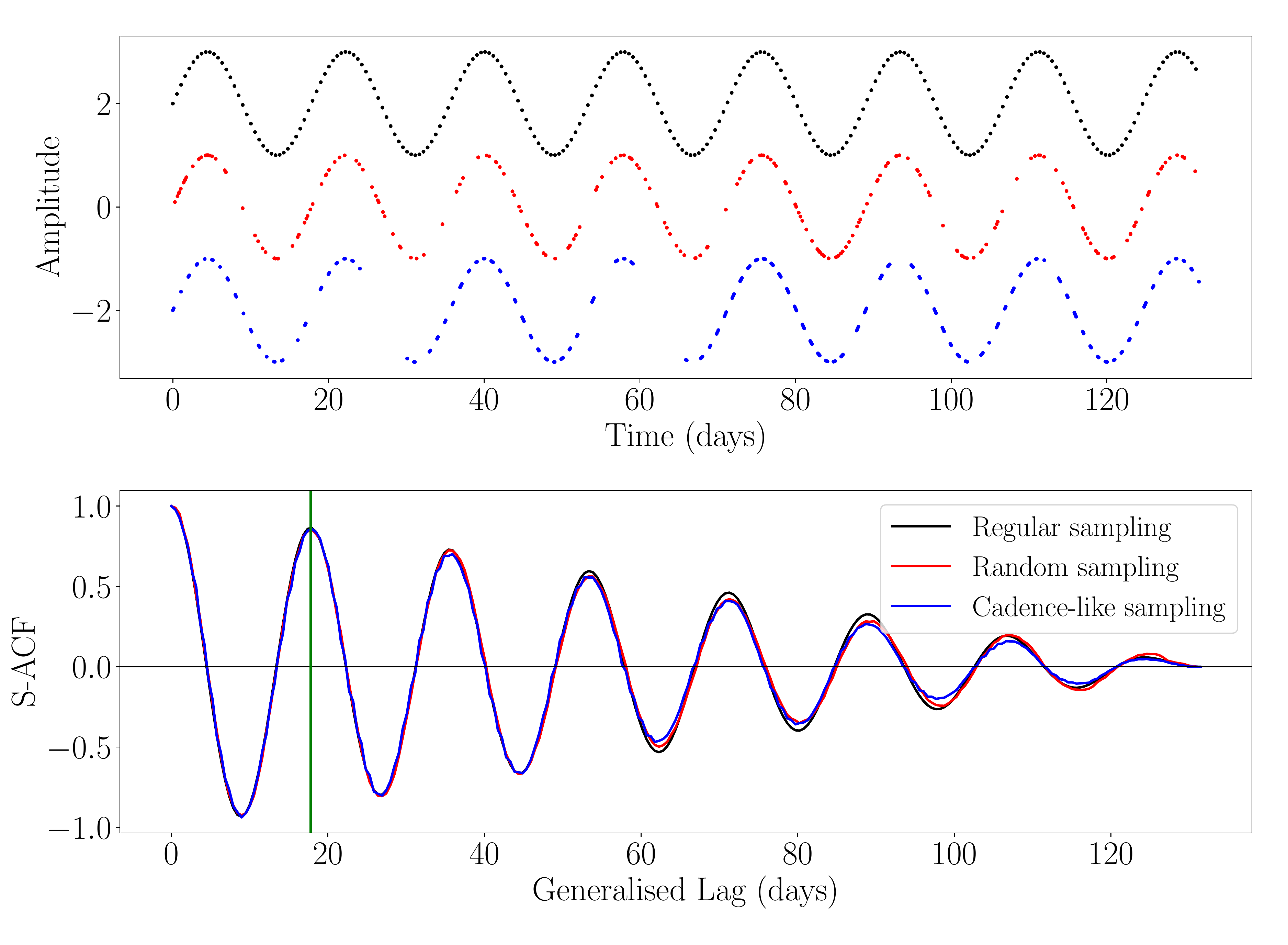}
	\caption{
    The top panel shows three time series with an underlying sine process ($17.8$ day period), sampled regularly (black), randomly (red) and with a cadence-like sampling that possesses additional larger gaps (blue). 
	All time label sets have the same cardinality of $|T_I|=250$. The bottom panel shows the selective autocorrelation functions (\sacf{}) of the above time series. A vertical green line is plotted at the period of the signal (17.8 days) in generalised lag.
	}
	\label{f:sin_process}
\end{figure}

\begin{figure}
  \includegraphics[width=\linewidth]{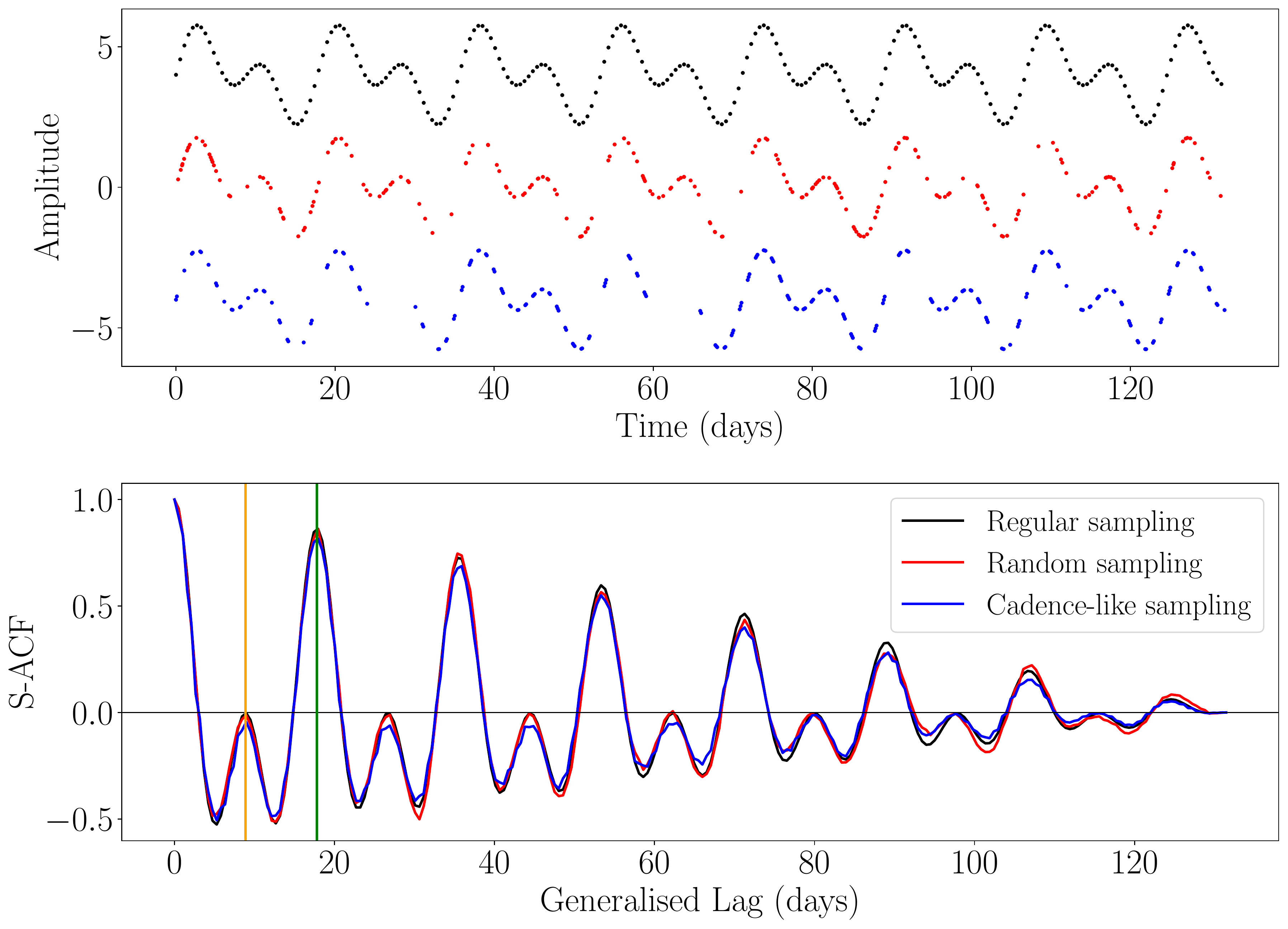}
	\caption{As Figure~\ref{f:sin_process} but for a time series with an underlying process described by the sum of two sine functions with 8.9 and 17.8 day periods. Vertical orange and green lines are plotted at the periods of the signal (8.9 and 17.8 days respectively) in generalised lag.
	}
	\label{f:two_sin_process}
\end{figure}

\begin{figure}
  \includegraphics[width=\linewidth]{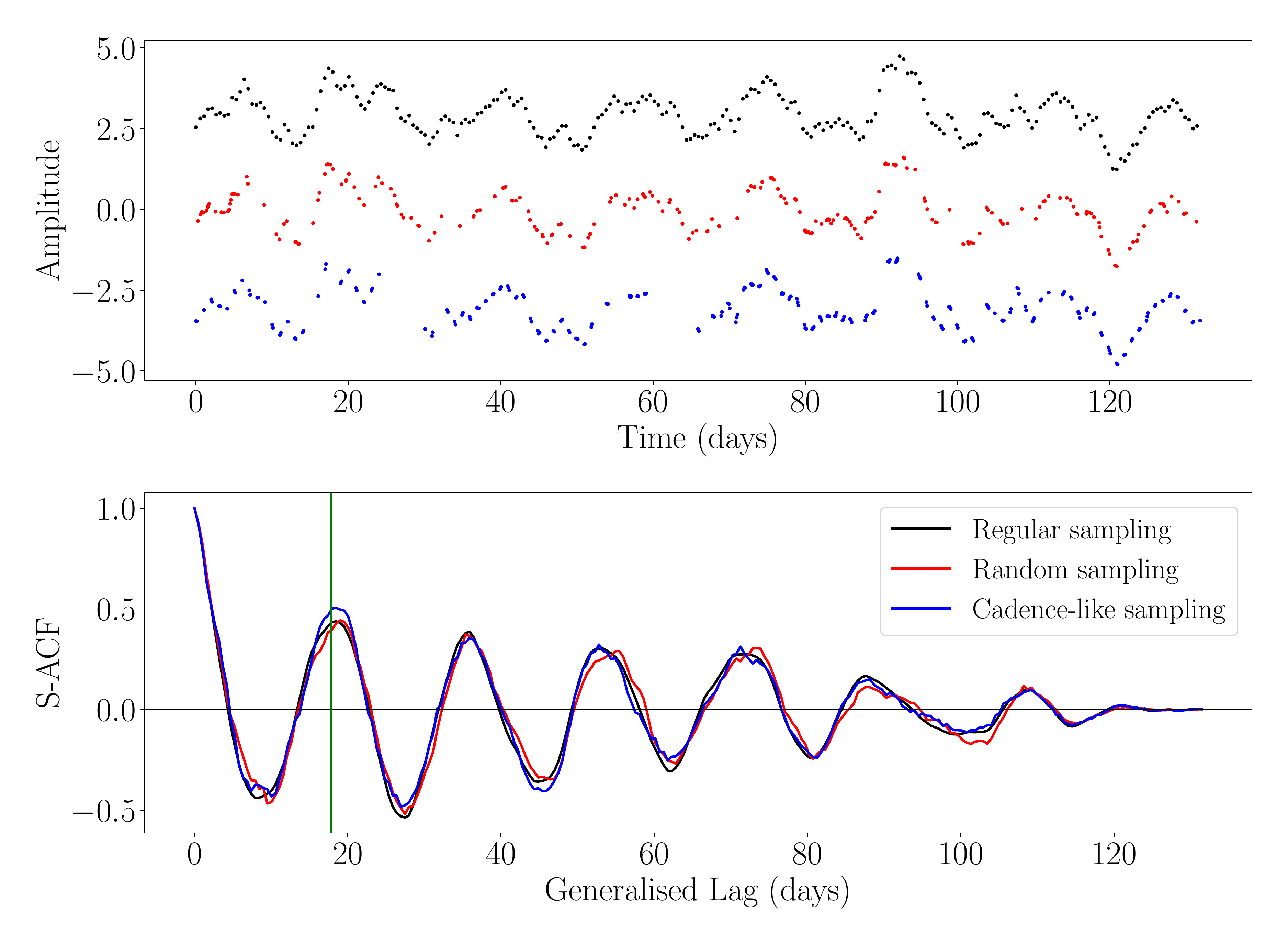}
\caption{As Figure~\ref{f:sin_process} but for a time series with an underlying process described by the sum of a comparable amplitude sine function and stochastic Gaussian process. A vertical green line is plotted at the period of the deterministic component of the signal (17.8 days) in generalised lag.
	}
	\label{f:GP_driven_process}
\end{figure}

\subsection{The weight function \texorpdfstring{$\hat{W}$}{W}}

We define a weight function $\hat{W}$ to be a function ${\hat{W} : [0,\infty) \to [0,1]}$ with $\hat{W}(0)\equiv 1$. We will interpret the weight function as a function that assigns time differences $\delta t \geq 0$ a weight within the interval $[0,1]$. Specifically we see from Equation \ref{G-ACF} that the weight function is used to assign a weight to the difference between the argument and the value of the selection function and is thereby a statement about the quality of the `selection'. Every fixed point of the selection function $\hat{S}\left(t_i + \hat{k}\right)=t_i + \hat{k}$ will therefore lead to a term in the \sacf{} with weight equal to one because of the requirement that $\hat{W}(0)\equiv 1$.\\

There are many choices for possible weight functions, but we have to make sure that the condition $\hat{W}(0)\equiv 1$ is observed since this is an important property (see Appendices \ref{proofofthereduction} and \ref{Proof-of-property-2}) and turns out to be a natural condition to ask for. Additionally, it would be natural for the weight function to be a monotonically decreasing function tending towards zero, since this reflects the interpretation that terms that involve time series values at similar points in time should be preferred. There are infinitely many such functions, including an exponential function or one half of a Gaussian distribution -- however a rational function is simpler and arguably more natural. 

The Python implementation published with this work supports several different weight functions. Simple tests indicate that different weight functions, that fit the above criteria, do not lead to clearly significant changes in the accuracy of the method. Finding the optimal weight function is beyond the scope of this paper.

We propose the following weight function 
\begin{equation}\label{weight-function-eq}
 \hat{W}(\delta t) = \frac{1}{1+ \alpha \delta t} , \quad \alpha > 0,\, \delta t \geq 0
\end{equation}
where $\alpha$ is the inverse of the characteristic scale parameter of the time series labels, e.g. one may choose $\alpha = 1/\langle T_I\rangle$. The $\delta t$ represents a generic time difference and does not have any interpretation as a sampling constant. 

Naively one may expect that the \sacf{} will depend on the scale of the time labels since we are free to re-scale time labels arbitrarily, but we know that the correlation between the different points in time of a process should not depend on the overall time scale. The use of the inverse scale parameter cancels out any re-scaling of the time labels since it will re-scale in the inverse fashion. This is the simplest continuous weight function that fits the above criteria and is also the most efficient for explicit calculations.

It is possible to consider a discrete weight function that satisfies $\hat{W}(0)=1$ but is zero in all other cases, thus discarding all terms that do not have matching shifted time labels and hence eliminating the selection function from the definition. However in the case of irregularly sampled time series such a discrete weight function may eliminate the majority of the terms contributing to the \sacf{} for a given lag and even almost matching terms would not be considered. 
This method is best suited to the case of an almost regular sampling where a small percentage of values are missing from an otherwise regularly sampled time series. In this case most terms in the autocorrelation function will have a match when the lag corresponds to an integer multiple of the `regular' sampling constant and only a small number of terms without a matching time label need to be discarded.

\section{\sacf{} applied to synthetic and real data}
\label{sec:application}

\begin{figure*}
  \includegraphics[width=0.95\textwidth]{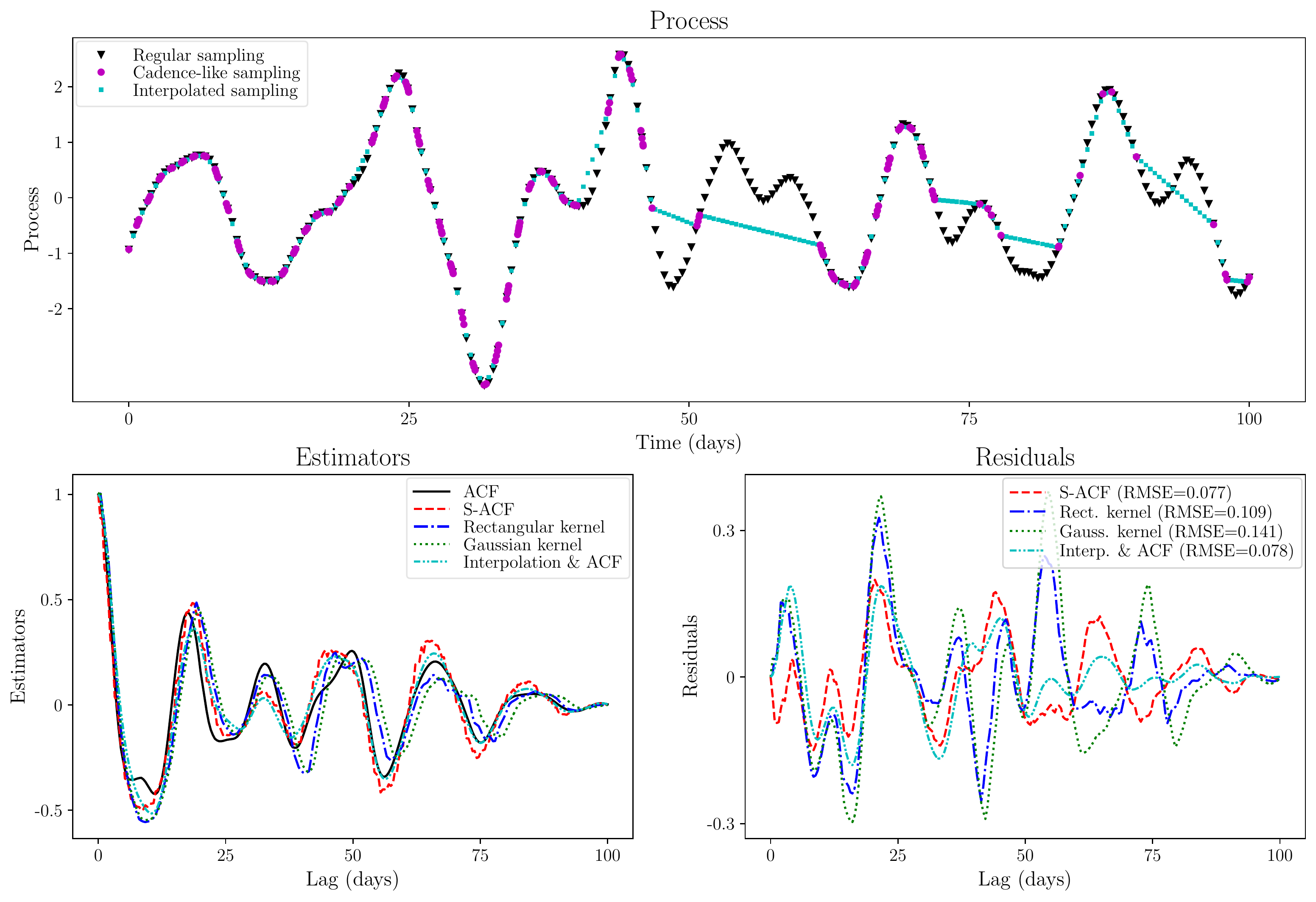}
	\caption{ \color{black}
    Top: three versions of a process comprising both a periodic component ($P = 17.8$ days) and a correlated noise component (characteristic timescale $l = 5$ days), which has an average sampling density of $2.5$ time labels per day and a signal-to-noise ratio of $5.0$.
    The three versions are a regular sampling of the process (black triangles), a cadence-like sampling (magenta circles) and an interpolation of the cadence-like process onto the regular sampling (cyan squares). 
    Bottom left: comparison of the different estimators of the autocorrelation function for this process. The standard estimator (black line) is based on the regularly sampled time series. The \sacf{} estimator (red dashed), rectangular kernel estimator (blue dash-dotted) and Gaussian kernel estimator (green dotted) are based on the cadence-like sampling of the process. 
    Finally, we use the interpolated regular sampling of the process to again apply the standard estimator (cyan dash-dot-dotted). Bottom right: residuals of
    the different estimators (\sacf{}, rectangular and Gaussian kernels, and interpolation) relative to the standard estimator (i.e. the ACF).
    Root-mean-square errors (RMSEs) are indicated in the legend.
	}
	\label{fig:estimators_comparison_good_data}
\end{figure*}

\begin{figure*}
  \includegraphics[width=0.95\textwidth]{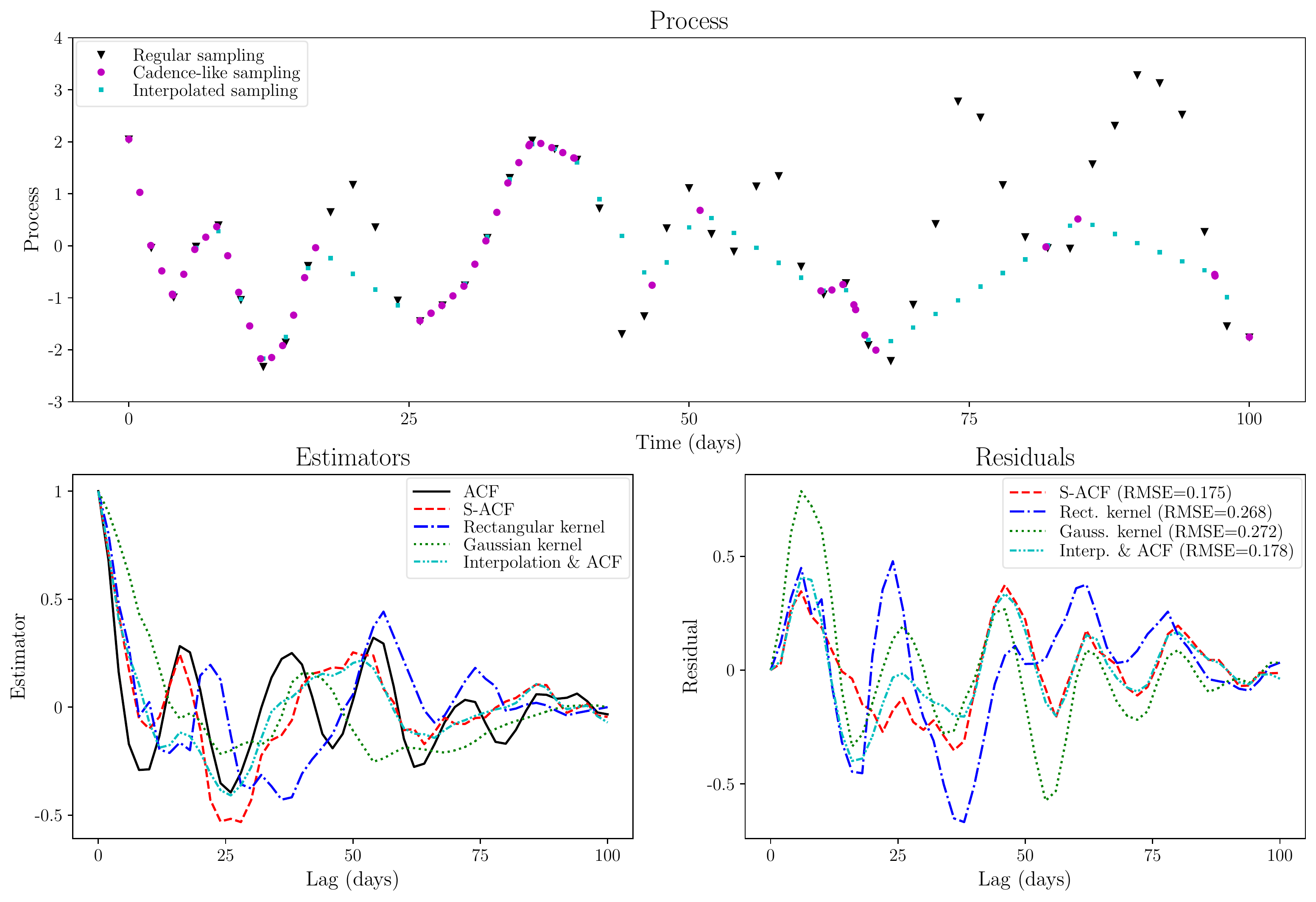}
	\caption{ \color{black}
 Same as Figure \ref{fig:estimators_comparison_good_data} for a process with a periodic component ($P = 17.8$ days) and a correlated noise component (characteristic timescale $l = 15$ days), an average sampling density of $0.5$ time labels per day and a signal-to-noise ratio of $0.1$.
	}
	\label{fig:estimators_comparison_bad_data}
\end{figure*}

\begin{figure*}
  \includegraphics[width=0.85\textwidth]{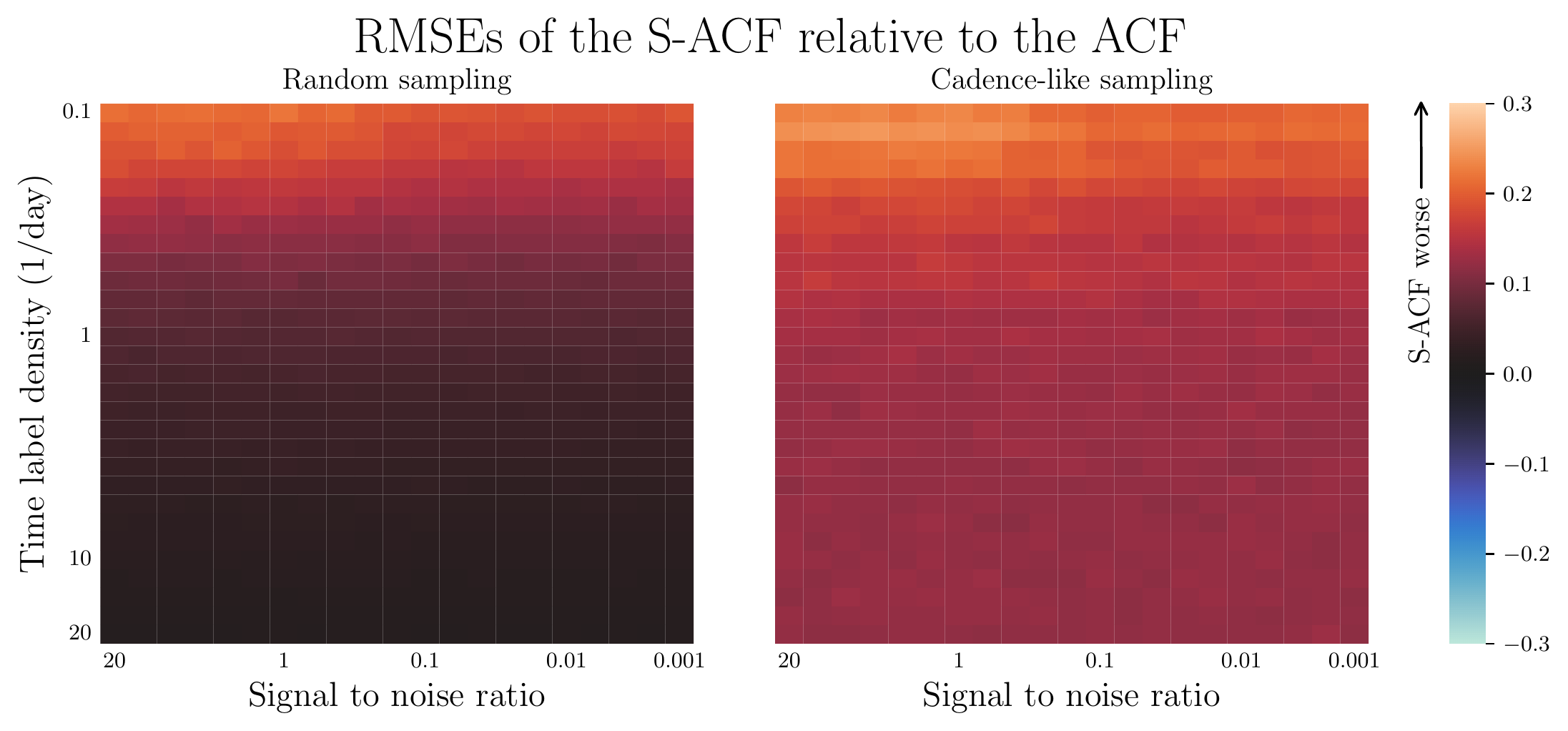}
	\caption{
    \color{black}
    Average RMSEs of the \sacf{} estimator for randomly sampled time series (left) and for time series with cadence-like sampling (right) relative to the standard estimator of the same process with regular sampling. 
    Each bin shows the average of many evaluations (totalling $140700$ processes per plot). Negative RMSEs are not possible and the colour scale is chosen to be consistent with Figures \ref{fig:heatmaps_random} and \ref{fig:heatmaps_cadence}, where differences between RMSEs are shown.
	}
	\label{fig:heatmaps_gacf_acf}
\end{figure*}

\begin{figure*}
  \includegraphics[width=0.95\textwidth]{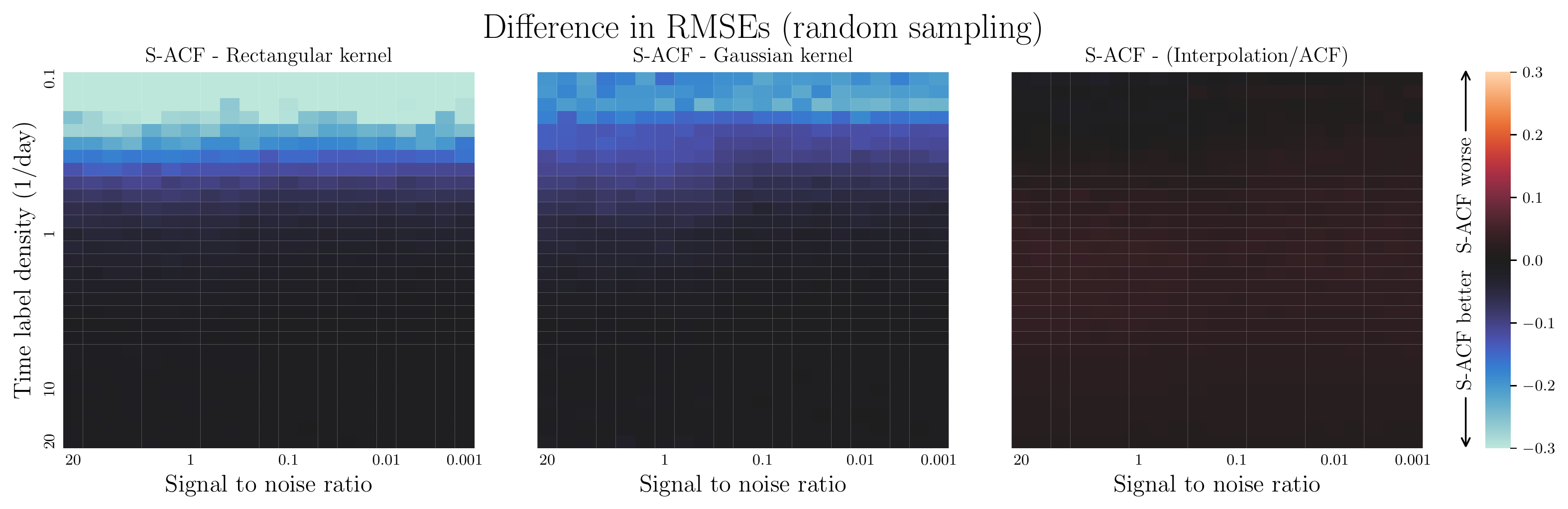}
	\caption{
 \color{black}
    Differences between RMSEs (relative to the standard estimator of the regularly sampled time series) using randomly sampled time series for the \sacf{} estimator vs. the rectangular kernel estimator (left), the Gaussian kernel estimator (centre) and the standard estimator on the interpolation of the randomly sampled time series (right). These are comparable to the RMSEs of the left panel of Figure \ref{fig:heatmaps_gacf_acf}.
    Red indicates a larger RMSE of the \sacf{} estimator, blue indicates a larger RMSE of the respective other estimator, and black indicates comparable RMSEs.
	}
	\label{fig:heatmaps_random}
\end{figure*}

\begin{figure*}
  \includegraphics[width=0.95\textwidth]{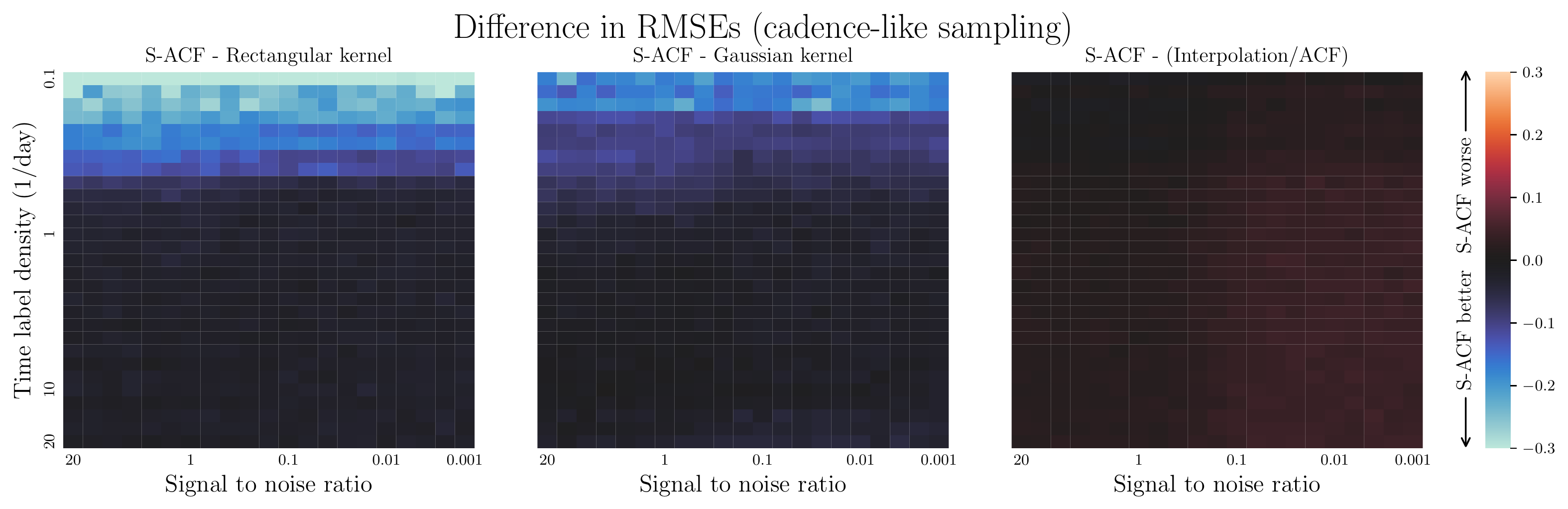}
	\caption{
    \color{black}
    Same as Figure \ref{fig:heatmaps_random} but for time series with a cadence-like sampling.
	}
	\label{fig:heatmaps_cadence}
\end{figure*}

\begin{figure*}
	\includegraphics[width=0.95\linewidth]{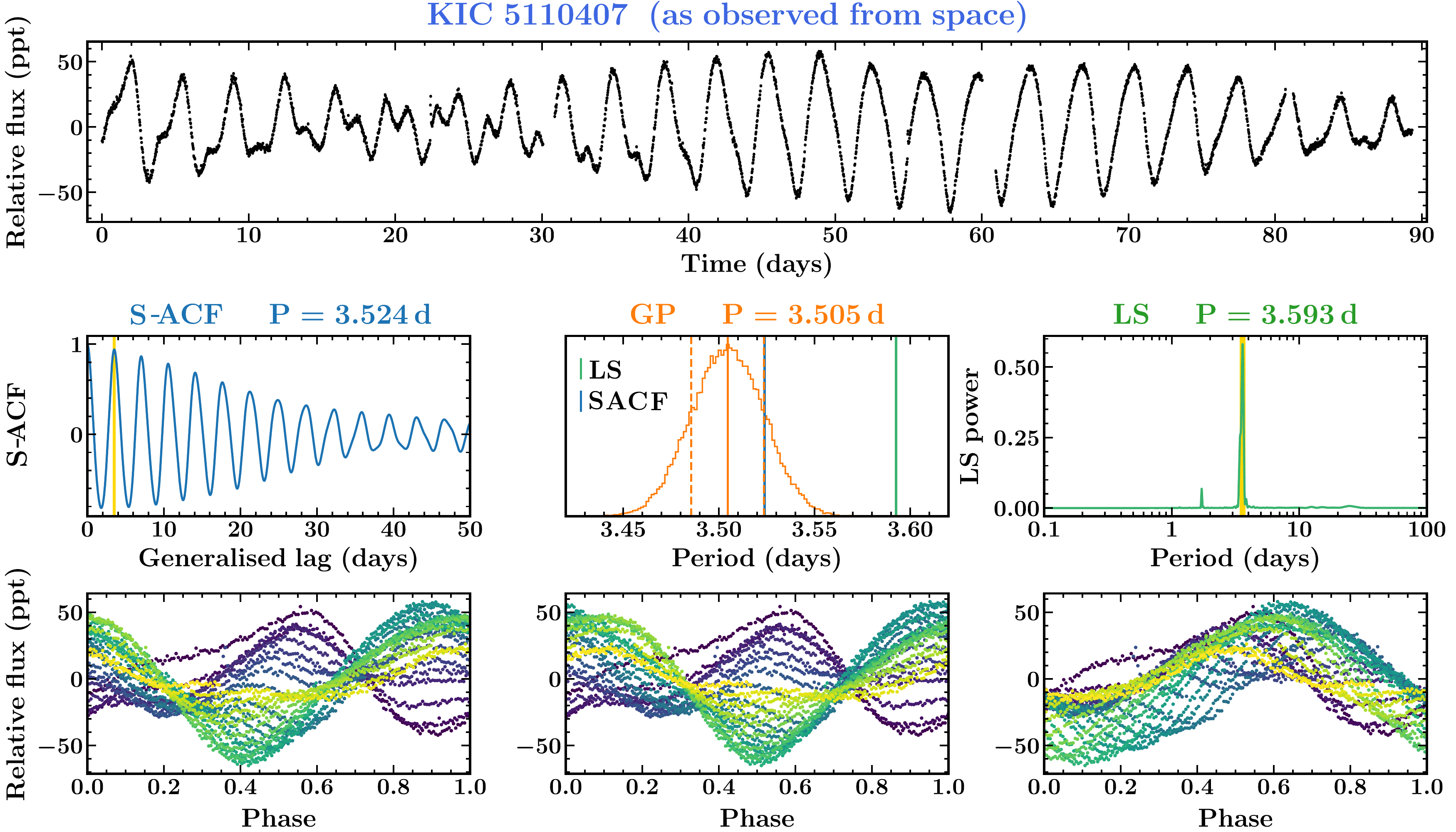}
    \caption{
    Rotation period estimates for the spotted star \kicID{} from \sacf{}, Gaussian process (GP) regression, and Lomb-Scargle (LS) periodogram. \emph{Top panel}: the system's quarter 7 \kepler{} light curve.
    \emph{Middle left}: Selective autocorrelation function (blue) with the identified period highlighted (yellow). 
    \emph{Middle centre}: GP posterior period distribution (orange) with the median and 1\,$\sigma$ uncertainties highlighted (solid and dashed orange lines). For comparison, the \sacf{} and LS periods are also shown (blue and green solid lines, respectively).
    \emph{Middle right}: LS periodogram (green) with the identified period highlighted (yellow). 
    \emph{Bottom row}: The \kepler{} light curve phase-folded on the corresponding method's period (\sacf{}, GP and LS; left-to-right) and coloured from the beginning (blue) to the end (yellow) of the observations.
    }
    \label{fig:P_comp_full}
\end{figure*}

\begin{figure*}
	\includegraphics[width=0.95\linewidth]{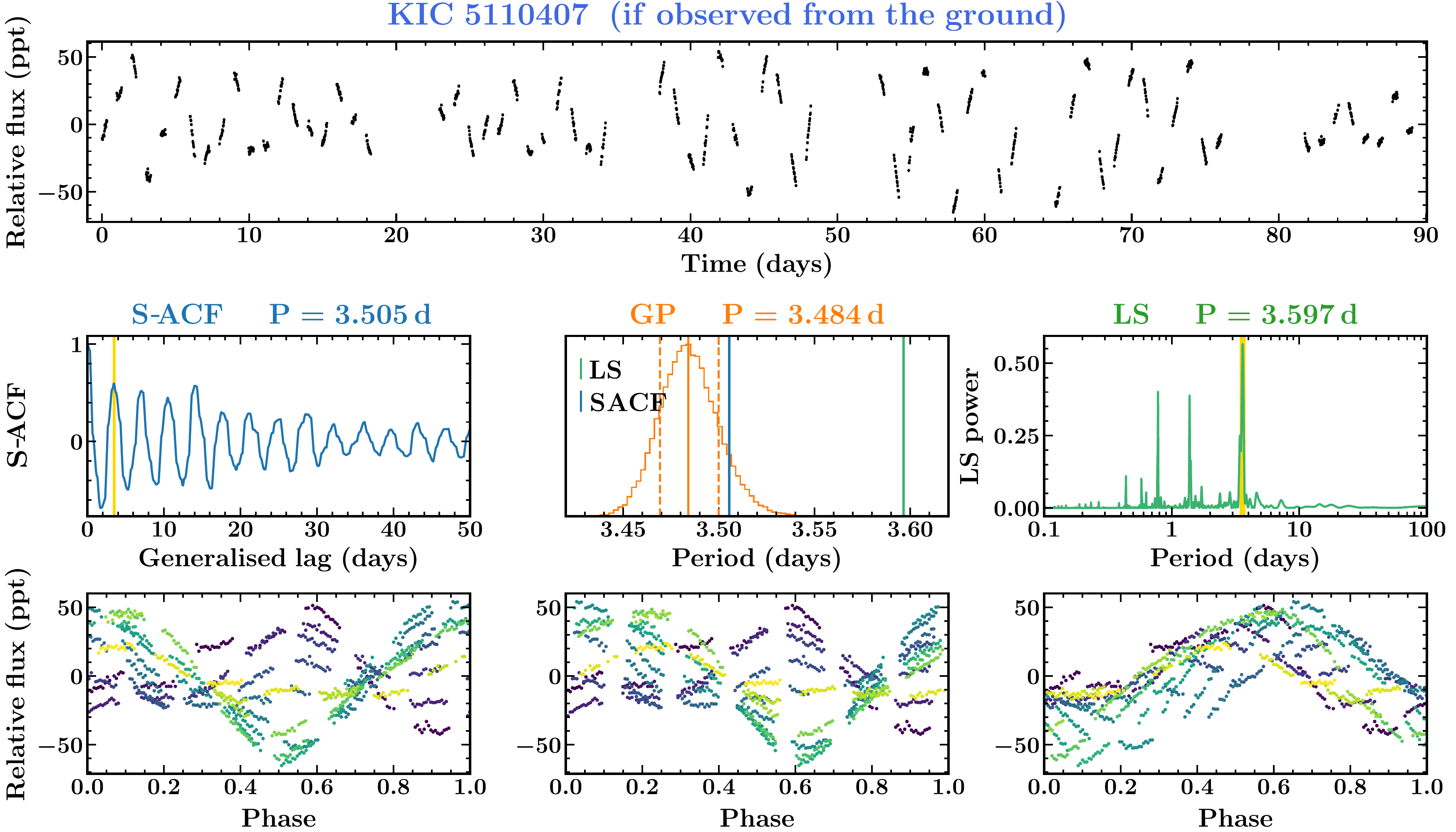}
    \caption{
    Same as Figure \ref{fig:P_comp_full} but simulating \kicID{} being observed from the ground (i.e. observations during night time only with additional gaps from bad weather).
    }
    \label{fig:P_comp_gappy}
\end{figure*}

\begin{figure*}
  \includegraphics[width=0.95\textwidth]{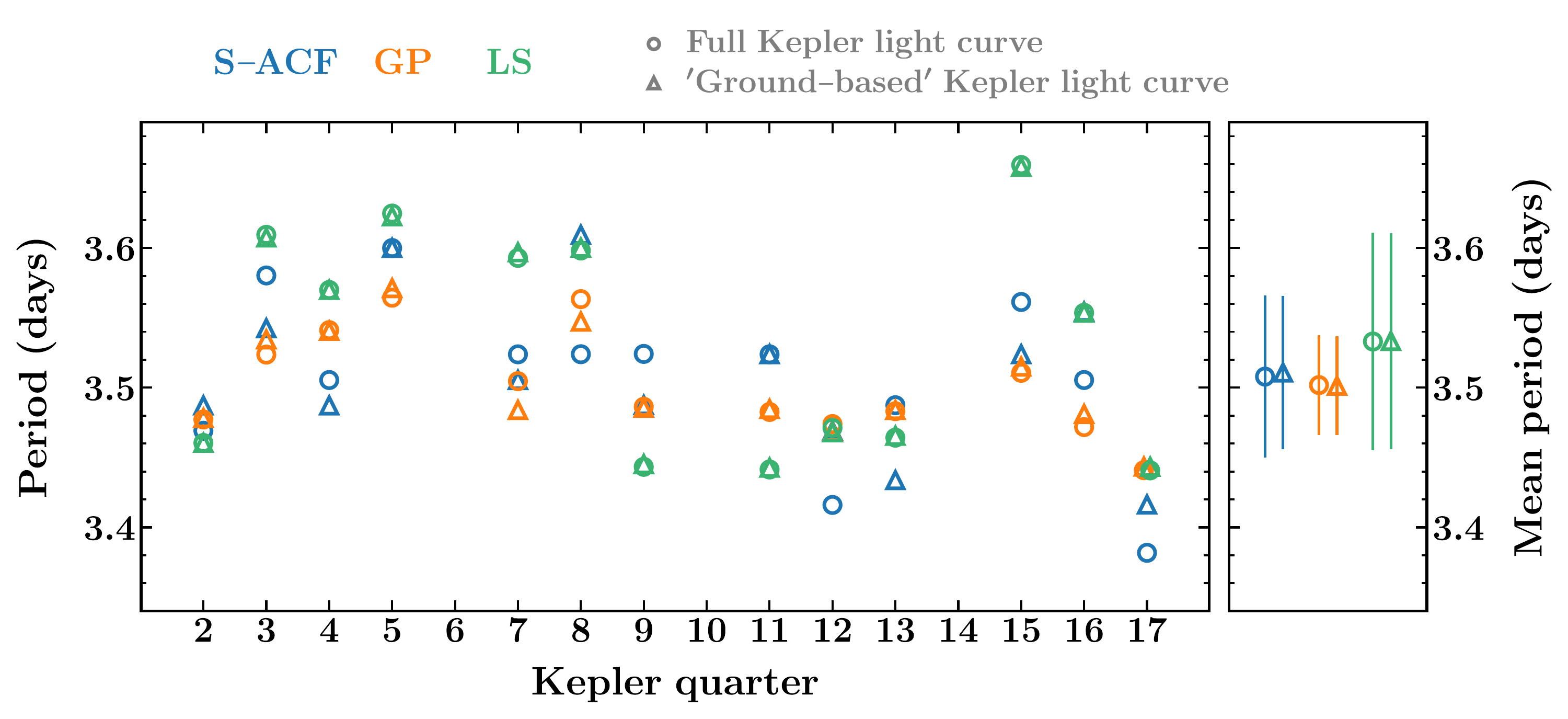}
	\caption{Rotation period estimates for \kicID{} from \sacf{} (blue), GP (orange) and LS (green) for all quarters with \kepler{} data. Circles show the period estimates from the full \kepler{} light curve and triangles from the `ground-based' version of the light curve.
	The \sacf{} periods typically agree better with the GP periods than with LS, especially in cases where there is evolution in the signal shape (e.g. quarters 7 and 15).
	The right hand panel shows the mean and standard deviation of the period estimates across all quarters. 
	The mean \sacf{} periods from both the full and `ground-based' versions of the light curves agree with each other, and with the values from both the GP and LS methods.
	The scatter in the \sacf{} periods across quarters is smaller than the scatter in LS periods but 
	larger than the scatter in GP periods.
	}
	\label{fig:quarter_comp}
\end{figure*}

\subsection{Synthetic data: simple sinusoidal time series}
\label{sec:application-synthetic}

We created a periodic sinusoidal signal. To investigate the impact of the temporal sampling on the \sacf{} we considered three different examples: (i) regularly sampled, (ii) randomly sampled, and (iii) cadence-like sampling with gaps. This third time series seeks to simulate the observing strategy of ground-based astronomical surveys, i.e. data during nighttime, gaps during daytime, and sporadic additional gaps due to bad weather. For ease of comparison, each time series contains the same number of data points ($|T_I|=250$) and they differ only in the temporal distribution of the data points.

These three time series are shown in the top panel of Figure \ref{f:sin_process}. The \sacf{} of the regularly sampled time series is identical to the ACF, as expected due to their definitions. The \sacf{} of the random and cadence-like samplings are similar to the ACF, but with small differences due to the data gaps and corresponding loss of information. These differences depend on the exact position and size of gaps within the data. 

In Figure \ref{f:two_sin_process} we consider the sum of two sine functions with the same three temporal samplings described above. Similar behaviour is seen in both the two-sine and single sine function examples, i.e. both the random and cadence-like sampling cases display modest deviations from the regularly sampled case but overall comparable autocorrelation functions.

\subsection{Synthetic data: more complex time series}
\label{sec:application-synthetic-GP}
In order to 
investigate
the efficacy of the \sacf{} on more realistic data sets, we generated a periodic signal with a large stochastic noise component. 
The periodic signal was again a sine function with a period of 17.8 days. 
The stochastic component was drawn from a Gaussian process (GP) using a simple harmonic oscillator (SHO) kernel (with quality factor $Q=1/3$ and characteristic timescale $\rho = 5$ days), as implemented in the {\tt celerite2} Python package \citep{Foreman-Mackey17,Foreman-Mackey18}.
The amplitude of the sinusoidal and stochastic components were comparable.

The same three temporal samplings were used as in Section \ref{sec:application-synthetic}, and the resulting time series and corresponding \sacf{}s are shown in Figure \ref{f:GP_driven_process}.
The \sacf{} displays a prominent peak 
corresponding to
the period of the sinusoidal component,
although the exact position of this peak 
will be moderately
affected by the large noise component, 
as expected. Despite the periodic and noise components having comparable amplitudes,
the \sacf{} is able to accurately recover a clear periodic signal 
in all three sampling cases.

{ \color{black}
\subsection{Synthetic data, quantitative analysis: comparison to kernel estimators and interpolation}
\label{sec:application-comparison-kernel-interpolation}

We want to quantitatively compare the performance of the \sacf{} to that of the standard autocorrelation estimator, as well as to several other methods. To do this, we assume that the standard autocorrelation estimator of a regular sampling of a process is close to the true autocorrelation function and thus we measure the other estimators relative to this function. We focus on comparing the estimators directly without determining the periods of the process. 

We consider the following methods: \sacf{}, the rectangular and Gaussian kernel estimators (as implemented in \citet{pastas-kernel-methods}), and the standard autocorrelation estimator following a linear interpolation of the irregularly sampled time series onto a regular sampling.

Throughout this section we consider the same kind of process as in section \ref{sec:application-synthetic-GP}, i.e. processes that consist of a GP rotation signal with an additional GP noise component.
The GP rotation component has a period between $0.1$ - $50$ days, standard deviation $\sigma = 1$, quality factor $Q_0=5$ (with $dQ=1$) and fractional amplitude of the secondary mode relative to the primary mode $f=0.5$. The noise is again drawn from a GP using a simple harmonic oscillator (SHO) kernel, with quality factor $Q=1/3$ and characteristic timescale $\rho$ between $0.1$ - $50$ days. The period of the periodic component of the process and the noise time scale are drawn uniformly at random from the interval $0.1$ - $50$. The signal-to-noise ratios considered are between $0.001$ and $20$. The overall length of the processes is $100$ days.
 
From these processes we generate time series with sampling densities (or time label densities) varying between $0.1$ - $20$ time labels per day. The length of each time series is kept fixed at $100$ days. The distributions of the time labels can be a uniform random distribution or a cadence-like distributions with regular sampling during night-times and additional larger gaps (e.g. simulating nightly observations with periods of bad weather).

The random sampling is generated by selecting time labels uniformly at random until the given average time label density is reached. The cadence-like sampling is generated by placing regularly sampled time labels during the night time (considered to last 8 hours) around 10 (larger) gaps, where the edges of the gaps are chosen uniformly at random. A small number of time labels are then added, at random, around the gaps until the given average time label density is reached.

Figures \ref{fig:estimators_comparison_good_data} and \ref{fig:estimators_comparison_bad_data} illustrate how we compare the different estimators in the case of cadence-like sampling (the process of comparing the different estimators is equivalent for the random sampling). Figure \ref{fig:estimators_comparison_good_data} shows an `easy' time series with an average sampling density of $2.5$ time labels per day and a signal-to-noise ratio (S/N) of $5$. We then compute the estimators derived from these methods based on the cadence-like sampling of the process. Finally, we compute the residuals and root-mean-square errors (RMSE) of each estimator relative to the standard autocorrelation estimator of the regularly sampled time series (i.e. no gaps). Figure \ref{fig:estimators_comparison_bad_data} repeats this for a `hard' time series with an average sampling density of $0.5$ time labels per day and a S/N of $0.1$. It is worth noting that the discrete time series shown in Figure \ref{fig:estimators_comparison_bad_data} (top) appear ``smoother'' than the underlying continuous process, due to their low sampling density.

Using the RMSEs computed in this way we can quantitatively compare the \sacf{} to the standard estimator for time series with random and cadence-like sampling. In the case of regular sampling the \sacf{} reduces to the standard estimator (see Appendix \ref{proofofthereduction}), which makes this comparison trivial by design. Figure \ref{fig:heatmaps_gacf_acf} shows the average RMSEs of the \sacf{} for a wide range of parameters. The RMSEs are generally much lower in the case of random sampling. Moreover we can see that the signal-to-noise ratio has essentially no effect when comparing the \sacf{} directly to the standard estimator of the regularly sampled time series, without applying further methods to detect periodicity. For sampling densities below $0.2$ the RMSEs increase in both sampling cases.

In order to compare the \sacf{} to other estimators of irregularly sampled time series, we compare the RMSEs of the \sacf{}, relative to the standard estimator of the regularly sampled time series, to the equivalent RMSEs of the other methods for the same processes. This is equivalent to running many analyses as in Figure \ref{fig:estimators_comparison_good_data} and taking the differences between the RMSEs of each method and the RMSE of the \sacf{}. The differences between the RMSEs are always computed for the same time series and are averaged over a large number of different time series.
In Figures \ref{fig:heatmaps_random} and \ref{fig:heatmaps_cadence} we show the results for time series with random and cadence-like sampling, respectively.

The \sacf{} performs better than the two kernel methods in all areas of the parameter space, but especially when the time label density is very low ($\leqslant 0.2$ day$^{-1}$), and this effect is more pronounced in the case of random sampling.
The performance of the \sacf{} is very similar to the interpolation method in all areas of the parameter space for the processes considered here. In the regime of very low S/N ratios ($\leqslant 0.1$) the interpolation method slightly outperforms the \sacf{}, at least in terms of the RMSE value. We note, however, that while the RMSE is a useful indicator of an accurate estimator, we are often concerned about extracting periods from these estimator functions, e.g. from the location of the first peak (or weighted average of the first few peaks). Considering Figure \ref{fig:estimators_comparison_bad_data}, as an illustrative example, the bottom left panel shows that only the \sacf{} has a first peak in general agreement with the ACF, so any period extracted from these estimator functions would likely favour \sacf{} over the other methods, at least in this particular example. Further tests, using a robust period finding algorithm across the full parameter space, would be needed to ascertain whether \sacf{} or interpolation produces more accurate period estimates. This is beyond the scope of this work.

}

\subsection{Real data: the \kepler{} light curve of the spotted star \kicID{}}
\label{sec:application-kepler}

We wish to test the efficacy of the \sacf{} on real time series data and explore how the periods estimated from the \sacf{} compare to other commonly used period estimation techniques. 
We estimate periods from the \sacf{} by calculating a Fast Fourier Transform (FFT) of the first 3 peaks of the \sacf{}. Restricting the lag time used in the period estimation reduces the effect of signal shape evolution on the autocorrelation function at long lag times and correspondingly improves the accuracy of the period estimated.
Using a FFT to calculate the periodicity of the \sacf{} is possible as the \sacf{} is a continuous function by definition. We note that another method of extracting periodicity from the \sacf{} would be to calculate the position of the first peak in the \sacf{}, or calculating the positions of subsequent peaks in addition in order to refine this period estimate, such as the technique used in \citet{McQuillan13}.

While the \sacf{} is not restricted to astronomy, the standard ACF has been widely used to estimate the rotation periods of stars from time series photometry. Therefore, as an illustrative example, we selected a spotted star observed by \kepler{}, \kicID{} \citep[e.g.][]{Roettenbacher13}, and compare the period predictions of \sacf{} to two other techniques for rotation period estimation: Gaussian process (GP) regression  and Lomb-Scargle (LS) periodograms. Using a rotationally variable star for this comparison allows us to probe the efficacy of \sacf{} on time series that display evolution in the signal (phase) shape. While we focus on the period here, we note that other useful information, such as the evolution timescale, can also be extracted from the \sacf{}. 
Our approach to comparing \sacf{}, GP regression and LS periodograms follows \citet{Gillen20a} and we refer the reader to Section 3 of that paper for further details, but give a brief overview below of the GP and LS models used here.

The GP model is based on the {\tt celerite2} package \citep{Foreman-Mackey17,Foreman-Mackey18}, as implemented through the {\tt exoplanet} framework \citep{exoplanet:joss,
exoplanet:zenodo}, and uses the standard rotation kernel with an additional simple harmonic oscillator (SHO) kernel (with quality factor $Q=1/3$) to capture any non-periodic structure in the light curves. 
The posterior parameter space was explored via gradient-based Markov-chain Monte Carlo (MCMC) using the No U-Turn Sampler (NUTS), as available through {\tt exoplanet}, which in turn uses {\tt PyMC3} and {\tt theano} \citep{Hoffman14,exoplanet:arviz, exoplanet:pymc3, exoplanet:theano}.
For each quarter, we ran 5 independent chains of 5,000 tuning steps followed by 10,000 sampling steps. GP periods were taken as the median of the posterior period distribution.
It is worth noting that the GP model requires an initial period guess, in contrast to both \sacf{} and LS, for which we give the average of the \sacf{} and LS period estimates. The GP model is also sensitive to data not well captured by the chosen rotation kernel, such as stellar flares, which we account for by performing an initial maximum a posteriori fit, masking 3\,$\sigma$ outliers, and refitting.
For the LS model, we use the version available through the {\tt astropy} project \citep{astropy13,astropy18}. LS periods are estimated from the largest peak in the periodogram. Both the LS and \sacf{} models were run on the data without further processing such as flare masking. 

\kepler{} observed \kicID{} for almost four years spanning 13 of the 17 quarters. \kepler{} quarters typically last $\sim$90 days and have essentially continuous observations with a cadence of $\sim$30 mins.
The ACF has been successfully applied to such \kepler{} data \citep[e.g.][]{McQuillan13,McQuillan14} but, as noted, the ACF is not applicable to non-continuous data that cannot be accurately interpolated onto a regularly spaced time series grid, i.e. time series with large data gaps, such as ground-based photometry.
We therefore estimated the stellar rotation period of \kicID{} from two versions of its \kepler{} light curve: (i) the full \kepler{} light curve and (ii) the \kepler{} light curve as though it had been observed from the ground (i.e. with gaps during daytime and simulated `bad weather' events\footnote{All quarters had the same relative times masked. Nighttime was considered to last 8 hours of each 24 hour period and bad weather was simulated between the following times: 18.5--22.5, 34.5--37.5, 48.5--52.5, 62.5--64.5 and 76.5--81.5 days (relative to the start of each quarter).}).
Figure \ref{fig:P_comp_full} shows the results for the full \kepler{} light curve observed during quarter 7 and Figure \ref{fig:P_comp_gappy} shows the results for `ground-based' version of the light curve.
The \kepler{} data from this quarter shows moderate evolution throughout and displays both `double-dip' patterns (e.g. at $\sim$20 days) and sinusoidal modulation (e.g. at $\sim$40--80 days). 
For this quarter, therefore, the \sacf{} and GP periods agree best whereas the LS period prediction is slightly larger. This is the case for both the full and `ground-based' light curves. The better agreement between \sacf{} and GP is because they are more flexible than LS (i.e. they do not assume a rigid sinusoidal model), and hence are more applicable to such evolving time series.
The periods can be best compared in the middle centre panel of Figures \ref{fig:P_comp_full} and \ref{fig:P_comp_gappy} and by comparing the phase-folded light curves.

We performed the same analysis on each available quarter of \kepler{} data and compare the period predictions for \sacf{}, GP and LS across quarters in Figure \ref{fig:quarter_comp}. Across quarters, and for both the full and `ground-based' light curves, the \sacf{} and GP periods agree best overall. The LS predictions agree well for some quarters, mainly those which show sinusoidal modulation, but less well for those that show evolving modulation patterns, which results in a larger scatter and correspondingly larger uncertainties on the mean rotation period prediction compared to \sacf{} or GP.
The mean periods and standard deviations across quarters are: 
\sacf{} = 3.51 $\pm$ 0.06 and 3.51 $\pm$ 0.06 days for the full and `ground-based' light curves, respectively; 
GP = 3.50 $\pm$ 0.04 and 3.50 $\pm$ 0.04 days; and 
LS = 3.53 $\pm$ 0.08 and 3.53 $\pm$ 0.08 days. 
We note that \citet{Roettenbacher13} estimate a rotation period for \kicID{} through  light-curve inversion of 3.4693 days, which agrees to within 1$\sigma$ for all three methods.

This comparison between the \sacf{} and the GP and LS methods, for both continuous and irregularly sampled time series, illustrates the validity of the \sacf{} for such applications. Furthermore, as the \sacf{} is a very general approach with minimal assumptions about the process, it can be applied to time series data of essentially any form, without the need to adapt the kind of model chosen (in the case of GP) or assume a rigid sinusoidal model (in the case of LS). 
The \sacf{} method took approximately 0.6 seconds on the quarter 7 \kicID{} light curve (4,117 data points) using a single laptop core\footnote{The run time of the \sacf{} is dependent on both the number of data points and the number of lag time steps.}.
The GP regression took $\sim$8.3 seconds for the maximum a posteriori fit (and $\sim$7 minutes for the MCMC), while the LS periodogram took approximately 0.01 seconds to run on the same laptop core. 
Each method is performing different calculations, and the GP MCMC method additionally provides a period uncertainty, so their respective times are simply included here for completeness and general interest. 
The \sacf{} is a powerful and efficient approach to extract periodicity, quasi-periodicity and short-term self-similarity from time series data in general, and especially data for which the true functional form is unknown.

Furthermore, we note that the \sacf{} has been successfully applied to ground-based data from the Next-Generation Transit Survey (NGTS; \citealt{Wheatley18}) to extract various kinds of stellar variability, including rotation, pulsations and eclipsing binaries \citep{Gillen20a,Briegal22}.

\subsection{A note on aliasing}
\label{sec:aliasing}

In the case of cadence-like sampling (e.g. ground-based astronomical surveys), which possess a (mostly) fixed periodicity of sampling gaps, alias signals can appear in the \sacf{} due to the missing information in between well-sampled clusters of data.
These aliases can be easily identified since their periodicity will be equal to the periodicity of the data clusters and their amplitude will be proportional to the relative size of the gaps between clusters \citep[see e.g.][]{Briegal22}. This effect will not be relevant for most applications unless the structure (or period) of interest is comparable to the structure of the sampling clusters (i.e. sidereal day for ground-based astronomical surveys).
We suspect that it may be possible to reduce this modest effect by generalising the normalisation to a function that depends on both the lag and time label density.
However, a full removal of these aliases will likely not be possible since gaps imply missing information that cannot be restored without additional information or assumptions. We note, however, that these effects are small if there are sufficient data points per period of the process.

\section{Conclusions}
\label{sec:conclusions}

The \sacf{}, or selective autocorrelation function estimator, is a new and versatile definition that can reliably and efficiently extract - amongst others - periodicity and signal shape information from any time series, virtually independent of the time series sampling and only assuming the smoothness of the underlying process. We show that the standard estimator of the autocorrelation function can be generalised and applied to irregularly sampled time series by generalising the lag to a real variable and introducing both selection and weight functions. We show that the \sacf{} reduces to the standard estimator for regularly sampled time series and possesses the property of maximal correlation at zero lag.

The \sacf{}s derived from both simple and more complex synthetic time series with different samplings (regular, random and `cadence-like') agree well, however there are small deviations due to the data gaps and corresponding loss of information.
We calculate the root-mean-square error (RMSE) of the \sacf{} of irregularly sampled synthetic processes, relative to the standard estimator of the ACF of the same process with a regular sampling. The RMSEs are calculated for a large number of processes spanning a wide range of time label densities and signal-to-noise ratios. The RMSEs of the \sacf{} are then compared to the equivalent RMSEs of other methods that aim to estimate the true ACF, including Gaussian and rectangular kernel estimators and a combination of linear interpolation and the standard estimator. The RMSEs of the \sacf{} increase significantly at low time label densities ($< 1$ day$^{-1}$) and the same effect can be seen with the other methods. The RMSEs of the \sacf{} have essentially no dependency on the signal-to-noise ratio. For the processes considered here, the \sacf{} performs better than the two kernel methods (most notably at low time label densities) and comparable to the interpolation method (although we note that the interpolation method performs slightly better at low signal-to-noise ratios, which may be due to properties of the GP noise that dominates the process in this regime). At high time label densities and modest-to-high S/N all considered methods perform well and are very close to the standard estimator of the ACF.

We compare the period predictions of \sacf{} to those from GP regression and LS periodograms by extracting rotation periods for the spotted star, \kicID{}. The \sacf{} and GP periods typically agree best across the different \kepler{} quarters, with LS periods being comparable in quarters with mainly sinusoidal modulation but more discrepant for quarters displaying more complex or evolving patterns. All three methods achieve consistent mean periods and uncertainties. 

There are a wide range of potential applications for the \sacf{}, not only within astronomy but also in other quantitative sciences where irregularly sampled time series occur, such as economics, finance, climatology, geology, biology and others. 

The Python implementation used in this work is available open-source under the MIT license at 
\href{https://www.github.com/joshbriegal/sacf}{github.com/joshbriegal/sacf}, 
and additionally can be installed through \textsc{PyPI} using the command: \texttt{pip install sacf}.

\section*{Acknowledgements}

LTK would like to thank the Bridgwater bursary of the Faculty of Mathematics at the University of Cambridge for their financial support and the Battcock Centre, Trinity College Cambridge and Sidney-Sussex College Cambridge for their hospitality.
EG gratefully acknowledges support from the David and Claudia Harding Foundation in the form of a Winton Exoplanet Fellowship.
JTB acknowledges support from the Science and Technologies Facilities Council (STFC) as part of the Centre for Doctoral Training in Data Intensive Science.
The authors thank Dan Foreman-Mackey for his insightful comments during the refereeing, helping them to improve the paper.

\section*{Data Availability}

The \kepler{} data used in this article are available through the Mikulski Archive for Space Telescopes (MAST) portal system at the URL \href{https://mast.stsci.edu}{https://mast.stsci.edu}.

\bibliographystyle{mnras}
\bibliography{ref} %

\appendix

\section{Notation}
\label{notation}

Table \ref{t:notation} lists the notation used in this work.

\begin{table*}
\caption{A brief summary of the notation of the sets, functions and parameters used.}
\centering
\resizebox{0.6\textwidth}{!}{%
\begin{tabular}{@{}ll@{}}
\toprule
Symbol & Description \\ \midrule
$I$ & Finite index set of natural numbers \\
$i_\text{max} := \max(I)$ & The maximum of the index set \\
$X(t)$ & A continuous real process \\
$X_I(t)$ & A time series with index set $I$ \\
$X_i=X(t_i)$ & A time series value corresponding to the label $t_i$ \\
$X_I$ & The set of time series values \\
$T_I$ & The set of time labels \\
$\Delta t$ & A positive sampling constant \\
$\rho(k)$ & The standard estimator of the autocorrelation function (ACF) \\
$k$ & The integer lag of the standard estimator \\
$\langle T_I\rangle$ & The mean value of the time label set \\
$\langle X_I\rangle$ & The mean value of the time series value set \\
$N := \sum\limits_{i\in I} \left(X_i - \langle X_I\rangle\right)^2$ & The normalisation of the ACF/\sacf{}\\
$\hat\rho(\hat k)$ & The selective autocorrelation function estimator (\sacf) \\
$\hat k$ & The real generalised lag of the \sacf{}\\
$\hat W(\delta t)$ & The weight function of the \sacf{}\\
$\delta t$ & A generic positive time difference \\
$\hat S(t)$ & The selection function of the \sacf{}\\
$\alpha$ & A positive constant \\\bottomrule
\end{tabular}%
}
\label{t:notation}
\end{table*}

\section{Proof of the reduction of the \sacf{} to the standard estimator for regularly sampled time series}
\label{proofofthereduction}
From the definition of the \sacf{} (Equation \ref{G-ACF}), the selection function (Section \ref{selectionfunction}) and the property $\hat{W}(0)\equiv 1$ of the weight function, we can derive a consistency property of the \sacf{} for the case of regularly sampled time series, which is that the \sacf{} reduces to the standard estimator in this case.

If a time series is regularly sampled, then there exists a sampling constant $\Delta t$ that describes the time difference between any two neighbouring time labels $t_i$ and $t_{i+1}$ by $t_{i+1}-t_i = \Delta t$ with $i,i+1 \in I$. In this case we can compute the standard estimator as well as the \sacf{}. If we want to compare the two functions for the same time series then we can only compare their values on their common domain, which means that the generalised lag $\hat{k}$ of the \sacf{} has to satisfy the relation $\hat{k}=k\cdot\Delta t$ with respect to the lag $k$ of the standard estimator. 
We thus have to restrict the real generalised lag to the domain of the standard estimator, given by integer multiples of the sampling constant. 

Using the restriction on the generalised lag, all points in time of the form $t_i + \hat{k}$ will satisfy
\begin{equation}
t_i + \hat{k} = t_i + k\cdot\Delta t,
\end{equation}
but, since the time series is regularly sampled, adding multiples of the sampling constant will give another time label
\begin{equation}
\label{timelabeleq}
t_i + \hat{k} = t_{i+k} \in T_I.
\end{equation}
If we now apply the selection function to this equation we obtain that
\begin{equation}
\hat{S}\left(t_i + \hat{k}\right) = \hat{S}\left(t_{i+k}\right),
\end{equation}
but since the selection function -- by construction -- maps its argument to the closest time label, we can use the fact that time labels are fixed points of the selection function, meaning that 
\begin{equation}
\hat{S}\left(t_{i+k}\right) = t_{i+k}
\end{equation}
to arrive at the equation
\begin{equation}
\label{fixedpoint}
\hat{S}\left(t_i + \hat{k}\right) = t_{i+k}.
\end{equation}
This result will be central to reducing the \sacf{} to the definition of the standard estimator for regularly sampled time series. \\

\noindent
If we look at the definition of the \sacf{} (Equation \ref{G-ACF}), we see that it only differs from the standard estimator by the factor of the weight function and the insertions of the selection function. %
If we focus only on the last factor in Equation \ref{G-ACF}, we can directly apply Equations \ref{timelabeleq} and \ref{fixedpoint} to give
\begin{equation}
\hat{W}\left(|\hat{S}\left(t_i + \hat{k}\right)-\left(t_i + \hat{k}\right) |\right) = \hat{W}\left(|t_{i+k}-t_{i+k}|\right) = \hat{W}(0).
\end{equation}
But we defined the weight function to satisfy the property $\hat{W}(0)\equiv 1$ and thus we obtain
\begin{equation}
\hat{W}\left(|\hat{S}\left(t_i + \hat{k}\right)-\left(t_i + \hat{k}\right) |\right) = 1
\end{equation}
removing the weight factor from the definition of the \sacf{}. \\

\noindent
The second factor in the definition of the \sacf{} (Equation \ref{G-ACF})
\begin{equation}
X(\hat{S}(t_i+\hat{k})) - \langle X_I\rangle
\end{equation}
is also modified by the selection function. However we can apply Equation \ref{fixedpoint} and obtain
\begin{equation}
X(\hat{S}(t_i+\hat{k})) - \langle X_I\rangle = X_{i+k} - \langle X_I\rangle
\end{equation}
and thus we again recover the factor from the definition of the standard estimator. \\

\noindent
Since the first factor of the \sacf{} (Equation \ref{G-ACF}) is the same as in the standard estimator, the only further modification is the restriction 
\begin{equation}
t_i+\hat{k} \leq \max(T_I)
\end{equation}
on the sum. We can again use Equation \ref{timelabeleq} to obtain
\begin{equation}
t_{i+k} \leq \max(T_I).
\end{equation}
If we write this equation in terms of the indices we arrive at 
\begin{equation}
i+k \leq i_\text{max}
\end{equation}
which is equivalent to writing the upper limit of the sum on $i$ as $i_\text{max}-k$, as is the case in the definition of the standard estimator.\\ 

\noindent
The above proof shows that the \sacf{} reduces to the definition of the standard estimator if the time series is regularly sampled and we restrict the generalised lag to the domain of the standard estimator.

\section{The proof of \texorpdfstring{$\hat{\rho}(0)=1$}{Rho(0)=1}}
\label{Proof-of-property-2}

We prove that $\hat{\rho}(0)=1$ by considering the \sacf{} definition (Equation \ref{G-ACF}) at zero lag. For $\hat{k}=0$ we find
\begin{align}
  \hat{\rho}&(0) = \nonumber\\
  & \frac{1}{N} \mathlarger{\mathlarger{\sum}}\limits_{\substack{i\in I\\ t_i \leq \max(T_I) }} \Bigg[  \left(X(t_i) - \langle X_I\rangle\right) \times \left(X(\hat{S}(t_i)) - \langle X_I\rangle\right)
  \nonumber \\
 & \qquad\qquad\qquad\quad\times\,\hat{W}\left(|\hat{S}\left(t_i\right)-\left(t_i\right) |\right)\Bigg],
\end{align}
but since $\hat{S}(t_i)=t_i$ and $\hat{W}(0)=1$ we have
\begin{align}
  \hat{\rho}&(0) = 
  \frac{1}{N} \mathlarger{\mathlarger{\sum}}\limits_{\substack{i\in I\\ i \leq i_\text{max} }} \Bigg[  \left(X(t_i) - \langle X_I\rangle\right) \times \left(X(t_i) - \langle X_I\rangle\right)\Bigg].
\end{align}
Using the definition of the normalisation we arrive at the desired equation
\begin{equation}
 \hat{\rho}(0) =  \frac{N}{N} = 1.
\end{equation}

\bsp	%
\label{lastpage}
\end{document}